\documentclass[conference]{IEEEtran}
\IEEEoverridecommandlockouts

\usepackage{amsmath,amssymb,amsfonts,mathtools}
\usepackage{graphicx}
\usepackage{textcomp}
\usepackage[dvipsnames]{xcolor}
\usepackage[a4paper, total={184mm,239mm}]{geometry}
\usepackage[hidelinks]{hyperref}
\usepackage[capitalize]{cleveref}
\usepackage{multirow, makecell}

\crefname{figure}{Fig.}{Figs.}
\Crefname{figure}{Fig.}{Figs.}
\crefname{subfigure}{Fig.}{Figs.}
\Crefname{subfigure}{Fig.}{Figs.}

\usepackage{glossaries}

\usepackage{listings}
\lstdefinestyle{mystyle}{
    basicstyle=\linestretch{0.5}\ttfamily,
    keywordstyle=\bfseries\color{MidnightBlue},
    commentstyle=\itshape\color{OliveGreen},
    showstringspaces=false,
}
\crefname{lstlisting}{Listing}{Listings}
\Crefname{lstlisting}{Listing}{Listings}

\usepackage{siunitx}
\DeclareSIUnit[quantity-product = {}]\times{{$\times$}}

\usepackage{tabularx}
\usepackage{booktabs}

\usepackage{acro}
\DeclareAcronym{ai}{short=AI, long=artificial intelligence}
\DeclareAcronym{bfp}{short=BFP, long=block floating-point}
\DeclareAcronym{cc}{short=CC, long=core complex, long-plural=es}
\DeclareAcronym{csr}{short=CSR, long=Control and Status Register}
\DeclareAcronym{dnn}{short=DNN, long=deep neural network}
\DeclareAcronym{eew}{short=EEW, long=Effective Element Width}
\DeclareAcronym{emul}{short=EMUL, long=Effective LMUL}
\DeclareAcronym{fma}{short=FMA, long=fused multiply-add}
\DeclareAcronym{flen}{short=FLEN, long=floating-point register width}
\DeclareAcronym{fpu}{short=FPU, long=floating-point unit}
\DeclareAcronym{fp}{short=FP, long=floating-point}
\DeclareAcronym{isa}{short=ISA, long=instruction set architecture}
\DeclareAcronym{ipu}{short=IPU, long=integer processing unit}
\DeclareAcronym{matmul}{short=MatMul, long=matrix multiplication}
\DeclareAcronym{mx}{short=MX, long=Microscaling}
\DeclareAcronym{mxdp}{short=MX-DP, long=MX dot product}
\DeclareAcronym{mxdpa}{short=MX-DPA, long=MX Dot-Product-Accumulate}
\DeclareAcronym{mxmatmul}{short=MX-MatMul, long=MX~matrix multiplication}
\DeclareAcronym{lmul}{short=LMUL, long=Length Multiplier}
\DeclareAcronym{lsu}{short=LSU, long=load-store unit}
\DeclareAcronym{ocp}{short=OCP, long=Open Compute Project}
\DeclareAcronym{pe}{short=PE, long=processing element}
\DeclareAcronym{ppa}{short=PPA, long={power, performance, and area}}
\DeclareAcronym{rf}{short=RF, long=register file}
\DeclareAcronym{sew}{short=SEW, long=Selected Element Width}
\DeclareAcronym{ssr}{short=SSR, long=Stream Semantic Register}
\DeclareAcronym{rvv}{short=RVV, long=RISC-V Vector Extension}
\DeclareAcronym{vla}{short=VLA, long=vector-length agnostic}
\DeclareAcronym{vrf}{short=VRF, long=vector register file}
\DeclareAcronym{vau}{short=VAU, long=vector arithmetic unit}
\DeclareAcronym{vlsu}{short=VLSU, long=vector load-store unit}
\DeclareAcronym{vpe}{short=VPE, long=vector processing element}
\DeclareAcronym{vsldu}{short=VSLDU, long=vector slide unit}

\usepackage[backend=biber,style=ieee]{biblatex}
\usepackage{xspace}
\newcommand{\vmxdotp}{\textsc{Vmxdotp}\xspace}

\usepackage{threeparttable}
\usepackage{pifont}

\usepackage{microtype}
\usepackage{float}
\newfloat{customlisting}{t}{lol}
\floatname{customlisting}{Listing}

\usepackage[dvipsnames]{xcolor}
\usepackage{tikz}
\usetikzlibrary{shapes,arrows,shadows,positioning,arrows.meta,calc}

\newcommand*\circled[2]{\tikz[baseline=(char.base)]{
    \node[shape=circle,draw=none,{#2},text=white,fill={#2},inner sep=1pt] (char) {\small \textbf{#1}};
}}

\usepackage{fancyhdr}
\fancypagestyle{noticestyle}{
    \fancyhf{}
    \fancyfoot[L]{\small \color{gray}%
        \vspace*{-3em}
        
        This paper has been accepted for publication by the 2026 Design, Automation and Test in Europe Conference (DATE~'26). \\
        ~\\
        © 2026 IEEE.
        Personal use of this material is permitted.
        Permission from IEEE must be obtained for all other uses, in any current or future media, including reprinting/republishing this material for advertising or promotional purposes, creating new collective works, for resale or redistribution to servers or lists, or reuse of any copyrighted component of this work in other works.
    }
    
}

\begin{document}

\title{\vmxdotp: A RISC-V Vector ISA Extension for Efficient Microscaling (MX) Format Acceleration%
\thanks{
This work was supported in part by the Swiss State Secretariat for Education, Research, and Innovation~(SERI) under the SwissChips initiative, and by Huawei Zurich Research Center~(ZRC).}%
}

\author{%
\IEEEauthorblockN{%
    Max Wipfli\IEEEauthorrefmark{1},
    Gamze İslamoğlu\IEEEauthorrefmark{1},
    Navaneeth Kunhi Purayil\IEEEauthorrefmark{1},
    Angelo Garofalo\IEEEauthorrefmark{2}
    and
    Luca Benini\IEEEauthorrefmark{1}\IEEEauthorrefmark{2}
}%
\IEEEauthorblockA{
    \IEEEauthorrefmark{1}\textit{IIS, ETH Zurich}, Switzerland;
    \IEEEauthorrefmark{2}\textit{DEI, University of Bologna}, Italy
}
\IEEEauthorblockA{
    mwipfli@ethz.ch, \{%
        \href{mailto:gislamoglu@iis.ee.ethz.ch}{gislamoglu},%
        \href{mailto:nkunhi@iis.ee.ethz.ch}{nkunhi},%
        \href{mailto:lbenini@iis.ee.ethz.ch}{lbenini}%
    \}@iis.ee.ethz.ch,
    angelo.garofalo@unibo.it
}
}

\hypersetup{
    pdftitle={VMXDOTP: A RISC-V Vector ISA Extension for Efficient Microscaling (MX) Format Acceleration},
    pdfauthor={Max Wipfli, Gamze İslamoğlu, Navaneeth Kunhi Purayil, Angelo Garofalo, Luca Benini},
}

\maketitle
\thispagestyle{noticestyle}

\begin{abstract}

Compared to the first generation of deep neural networks, dominated by regular, compute-intensive kernels such as \acp{matmul}\acuse{mxmatmul} and convolutions, modern decoder-based transformers interleave attention, normalization, and data-dependent control flow.
This demands flexible accelerators, a requirement met by scalable, highly energy-efficient shared-L1-memory \ac{vpe} clusters.
Meanwhile, the ever-growing size and bandwidth needs of state-of-the-art models make reduced-precision formats increasingly attractive.
\ac{mx} data formats, based on \ac{bfp} representations, have emerged as a promising solution to reduce data volumes while preserving accuracy.
However, \ac{mx} semantics are poorly aligned with vector execution:
block scaling and multi-step mixed-precision operations break the regularity of vector pipelines, leading to underutilized compute resources and performance degradation.
To address these challenges, we propose \vmxdotp, a RISC-V Vector (\acuse{rvv}\acs{rvv}) 1.0 \ac{isa} extension for efficient \ac{mx} dot product execution, supporting MXFP8 and MXFP4 inputs, FP32 and BF16 accumulation, and software-defined block sizes.
A \vmxdotp-enhanced \ac{vpe} cluster achieves up to \qty{97}{\percent} utilization on \ac{mxmatmul}\acuse{matmul}.
Implemented in \qty{12}{\nano\meter} FinFET, it achieves up to 125~MXFP8-GFLOPS and 250~MXFP4-GFLOPS, with 843/1632~MXFP8/MXFP4-GFLOPS/W at \qty{1}{\giga\hertz}, \qty{0.8}{\volt}, and only \qty{7.2}{\percent} area overhead.
Our design yields up to 7.0$\times$ speedup and 4.9$\times$ energy efficiency with respect to software-emulated MXFP8-\ac{matmul}.
Compared with prior \ac{mx} engines, \vmxdotp supports variable block sizes, is up to 1.4$\times$ more area-efficient, and delivers up to 2.1$\times$ higher energy efficiency.

\end{abstract}

\section{Introduction}

The growing memory, bandwidth, and compute requirements of modern \ac{ai} workloads present significant challenges.
To address these, one effective approach is the use of narrow bit-width data formats, which significantly reduce storage and data movement costs while enabling more energy-efficient computation.
However, as bitwidths decrease, preserving model accuracy becomes increasingly challenging due to the reduced dynamic range and precision~\cite{rouhani2023shared}.

To alleviate this trade-off, block-scaled data formats have emerged as a compelling solution.
By associating a shared scale factor with a block of low-bitwidth elements, these formats preserve high dynamic range while retaining the benefits of a compact representation.
In particular, the recently proposed \acf{mx} formats~\cite{ocpmx} couple a block-level exponent to a vector of narrow \ac{fp} elements.
Standardized by the \acf{ocp} and supported by key industry players, \ac{mx}~formats have demonstrated high accuracy across a wide range of \ac{ai} workloads, often serving as a drop-in replacement for wider formats~\cite{rouhani2023microscaling}.

While the memory savings of \ac{mx}~formats are a direct consequence of their compact design, their computational benefits are often overlooked.
\ac{mx} quantization is frequently treated as a storage-only compression approach to alleviate memory bottlenecks, requiring decompression to wider formats before computation~\cite{gerogiannis2025deca,qcom_mx}.
Fully exploiting the computational efficiency of \ac{mx}~formats, however, requires native hardware support.
Recognizing this, both NVIDIA and AMD have recently added such support in their \emph{Blackwell} and \emph{CDNA 4} microarchitectures, respectively~\cite{nvidia_blackwell,amd_cdna4}.

The success of \ac{mx}~formats in specialized hardware has naturally led to growing interest in supporting them on more general-purpose, programmable architectures~\cite{islamoglu2025mxdotp}.
In particular, vector processors are a promising target as they combine data parallelism, programmability, and software portability.
These features have led to their adoption in mainstream \acp{isa}, notably through Arm SVE, SVE2, and the recently ratified \acf{rvv} 1.0.
Among these, the open-source \ac{rvv}, explicitly designed for high efficiency on data-parallel workloads pervasive in AI, offers a compelling framework for supporting emerging standards such as \ac{mx} formats.

However, the optimal path for integrating \ac{mx} support into \ac{rvv} is not yet clear.
To enable software emulation of \ac{mx}~operations, narrow \ac{fp} elements must be cast to wider formats for computation.
To this end, a set of vector conversion instructions is in the process of being standardized for \ac{rvv}~\cite{zvfofp8min,zvfofp4min}.
Although an essential first step, this approach treats \ac{mx}~formats purely as a storage or transport medium.
As our analysis will show, this fails to address the core computational bottlenecks and can leave substantial performance and efficiency gains on the table.

This paper argues that unlocking the full computational benefits of \ac{mx}~formats on vector processors requires a tightly integrated hardware approach.
To demonstrate this, we extend Spatz, an open-source \ac{vpe}~\cite{perotti2025spatz}, with a custom \ac{rvv} \ac{isa} extension that enables direct \ac{mx} dot product computation without prior decompression, and make the following contributions:

\begin{itemize}
    \item We implement and analyze \ac{rvv}~kernels for software-emulated \ac{mxmatmul}, identifying fundamental performance limitations that cannot be addressed using only the existing \ac{isa} and \ac{fp} conversion instructions.
    \item We propose \vmxdotp, a novel RVV \ac{isa} extension that provides native, single-instruction support for MXFP8 and MXFP4 dot products with accumulation in FP32 or BF16 and flexible, software-defined block sizes.
    \item We integrate \vmxdotp into the Spatz \ac{vpe} and implement the design in a \qty{12}{\nano\meter} FinFET technology, incurring an area overhead of \qty{12.6}{\percent} at the core level, and only \qty{7.2}{\percent} at the cluster level.
    \item We demonstrate up to 7.0$\times$ speedup and 4.9$\times$ higher energy efficiency for \ac{mxmatmul} compared to software emulation on the original Spatz processor.
\end{itemize}
\section{Background}

\subsection{Microscaling (MX) Formats}

The \ac{ocp} \acf{mx} specification~\cite{ocpmx} defines a class of \ac{bfp} formats.
Each \ac{mx} block contains $k$~elements sharing a single 8-bit exponent scale~(E8M0), increasing the dynamic range between blocks despite the compact representation.
The specification defines several concrete data formats, all with a block size of $k = 32$.
There are five formats with \ac{fp} elements (MXFP8\textsubscript{E5M2/E4M3}, MXFP6\textsubscript{E3M2/E2M3}, and MXFP4\textsubscript{E2M1}) as well as the MXINT8 format with 8-bit signed integer elements.

The fundamental operation on MX data is the dot product (\acuse{mxdp}\acs{mxdp}) between two \ac{mx}~blocks, $A$ and $B$, defined as:
\begin{equation}
    C=\mathrm{Dot}(A,B) = X(A) \cdot X(B) \cdot \sum_{i=1}^k P_i(A) \cdot P_i(B),
\end{equation}
where $X(A)$, $X(B)$ are the block scales, $P_i(A)$, $P_i(B)$ the individual elements, and the result $C$ \emph{should} be in FP32 format.

This work focuses on MXFP8 and MXFP4 formats.
We omit MXFP6 as its 6-bit elements are ill-suited to byte-oriented general-purpose processors, and exclude MXINT8, as it can be efficiently emulated using integer arithmetic~\cite{satyamurthy2024optimization}.

\subsection{RISC-V Vector Extension (RVV)}
The standardized \acf{rvv} adds a data-parallel programming model to the \ac{isa}.
Instructions are configured via the \texttt{vtype} \ac{csr}, which includes \ac{sew} to define operand widths.
\ac{rvv} supports widening and narrowing operations, where the result element width is twice or half the operand element width, respectively.
For these operations, the wider data type uses an \acf{eew} of $2 \times \text{\ac{sew}}$.
To increase register utilization, \ac{rvv} uses \acp{lmul} to combine multiple VLEN-bit registers into longer vectors.
To match the number of elements in all operands in mixed-width operations, an \ac{emul} of $2 \times \text{\ac{lmul}}$ is used for the wider operands.

\subsection{Spatz}
\label{sec:background_spatz}

Spatz~\cite{perotti2025spatz} is an open-source \ac{rvv} processor designed for energy efficiency and embedded applications.
It is coupled to a tiny 32-bit scalar integer core, forming a Spatz \ac{cc}.
The \ac{vpe} consists of a centralized \acf{vrf}, a controller, and three parallel functional units: the \ac{vau}, the \ac{vlsu}, and the \ac{vsldu}.
The \ac{vrf} hosts the 512-bit wide vector registers distributed across 4~banks, which provide three read ports and one write port (3R1W) each.
The \ac{vau} handles most computational instructions through its \ac{ipu} and four \acp{fpu}.
Spatz also handles scalar \ac{fp} operations, for which the controller hosts a separate \ac{rf} and a \ac{lsu}.

Two Spatz \acp{cc} form a cluster, sharing a \qty{128}{\kibi\byte} L1 scratchpad memory.
In total, the cluster can sustain 128~bits of integer or 512~bits of \ac{fp} operations per cycle.

Spatz supports most of the standard \ac{rvv} extension and two custom \ac{isa} extensions~\cite{bertaccini2024minifloats}:
\emph{MiniFloat-NN} adds full support for low-precision 16-bit (FP16, BF16) and 8-bit (FP8\textsubscript{E5M2/E4M3}) \ac{fp} formats.
\emph{ExSdotp} adds the \texttt{vfwdotp} instruction to compute an ``expanding sum of dot products.’’
It multiplies two \ac{fp} operand pairs and accumulates the results into a double-width destination register, effectively doubling the throughput compared to regular widening \ac{fma} instructions (\texttt{vfwmacc}).
\section{Software Emulation}
\label{sec:software_emulation}

\begin{figure}[t]
	\centering
    \vspace{-0.5em}
	\includegraphics{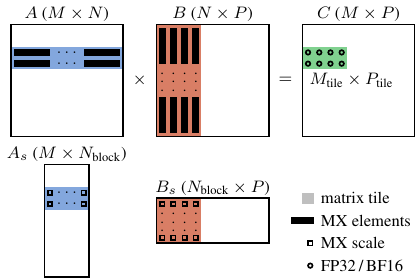}
    \vspace*{-0.3em}
	\caption{
		Overview of \ac{mxmatmul} with $M_\text{tile} \times P_\text{tile}$ output tiles.
		The separate scale matrices ($A_s$, $B_s$) have reduced inner dimensions of $N_\text{block} = N / k$.
	}
	\label{fig:mx_matmul}
\end{figure}

To show the limitations of emulating \ac{mx} operations fully in software, we implement \ac{rvv} kernels for MXFP8-\ac{matmul}.
As illustrated in \cref{fig:mx_matmul}, data is quantized into \ac{mx} blocks along the reduction axis (i.e. along rows for $A$, columns for $B$).
The E8M0 block scales are stored separately (in $A_s$, $B_s$) from the \ac{fp} elements (in $A$, $B$).
For accumulation, we consider both the specified FP32 format and the more compact BF16 format.

\subsection{FP8 Conversion Support}

The kernels require vector and scalar instructions to expand FP8 operands to 16~bits for further processing.
As this is not supported in standard \ac{rvv}, we use the \emph{MiniFloat-NN} extension~\cite{bertaccini2024minifloats} implemented by the baseline Spatz \ac{vpe}.

The proposed \emph{Zvfofp8min} standard extension~\cite{zvfofp8min} provides FP8-to-BF16 vector conversion instructions as an alternative, but lacks a scalar counterpart.

\subsection{Baseline: MXFP8-\ac{matmul} Kernel}

Our implementation uses an outer-product algorithm, which vectorizes computation along the output matrix's row dimension, avoiding often inefficient reduction instructions (\texttt{vfredusum.vs}).
We assume that all input matrices are stored in row-major order, which avoids inefficient strided 8-bit loads on $B$ and $B_s$.
The pseudocode in \cref{lst:kernel_baseline} illustrates the computation for a $1 \times P_\text{tile}$ output tile with FP32 accumulation.

\begin{customlisting}
\begin{lstlisting}[basicstyle=\footnotesize\ttfamily,
    caption={
        Baseline \ac{rvv} kernel for MXFP8-\ac{matmul} ($1 \times P_\text{tile}$ output tile) with FP32 accumulation.
        \vspace{1mm}
    },
    label={lst:kernel_baseline}]
size_t N, N_block = N / BLOCK_SIZE;
size_t P_tile = get_vlmax(SEW_32, LMUL_4);
fp8_t A[1][N];      e8m0_t As[1][N_block];
fp8_t B[N][P_tile]; e8m0_t Bs[N_block][P_tile];
float C[1][P_tile];
(*@\textbf{vsetvli}@*)(P_tile, SEW_32, LMUL_M4); v0..3 = (*@\textbf{vmv.v.i}@*)(0);
for (size_t block = 0; block < N_block; block++) {
   v4..7 = (*@\textbf{vmv.v.i}@*)(0);
   for (size_t elem = 0; elem < BLOCK_SIZE; elem++) {
     size_t idx = block * BLOCK_SIZE + elem;
     fp16_t a0 = (*@\textbf{fcvt.h.b}@*)(A[0][idx]);
     (*@\textbf{vsetvli}@*)(P_tile, SEW_8, LMUL_M1);
     v8 = (*@\textbf{vle8.v}@*)(B[idx][:]);
     v8..v9 = (*@\textbf{vfwcvt.f.f.v}@*)(v8);
     (*@\textbf{vsetvli}@*)(P_tile, SEW_16, LMUL_M2);
     v4..v7 = (*@\textbf{vfwmacc.vf}@*)(v4..v7, v8..v9, a0);
   }
   int as0 = As[0][block] - 127;      // remove bias
   v12 = (*@\textbf{vle8.v}@*)(Bs[block][:]);
   v12..v13 = (*@\textbf{vwcvtu.x.x.v}@*)(v12);
   (*@\textbf{vsetvli}@*)(P_tile, SEW_16, LMUL_M2);
   v16..19 = (*@\textbf{vwadd.vx}@*)(v12..v13, as0);
   v16..v19 = (*@\textbf{vsll.vi}@*)(v16..v19, 23);      // to FP32
   (*@\textbf{vsetvli}@*)(P_tile, SEW_32, LMUL_M4);
   v0..3 = (*@\textbf{vfmacc.vv}@*)(v0..v3, v4..v7, v16..v19);
}
c[0][:] = (*@\textbf{vse32.v}@*)(v0..v3);           // store result
\end{lstlisting}
\begin{tikzpicture}[overlay,x=0.7em,y=0.5em]
    \draw[RoyalBlue,thick,Bar-Bar] (1,21.2) -- (1,38.9)
        node[midway,shape=circle,black,inner sep=1pt,draw=none,fill=RoyalBlue,text=white]
        {\small \textbf{1}};
    \draw[Green,thick,Bar-Bar] (1,10.4) -- (1,20.9)
        node[midway,shape=circle,black,inner sep=1pt,draw=none,fill=Green,text=white]
        {\small \textbf{2}};
    \draw[BrickRed,thick,Bar-Bar] (1,7.0) -- (1,10.1)
        node[midway,shape=circle,black,inner sep=1pt,draw=none,fill=BrickRed,text=white]
        {\small \textbf{3}};
\end{tikzpicture}
\vspace{-1.8em}
\end{customlisting}

The kernel iterates through each \ac{mx}~block along the reduction dimension, where it performs three steps:
\circled{1}{RoyalBlue}~An inner loop iterates through the elements, producing an unscaled block dot product.
Each iteration loads an FP8 element from $A$ and the corresponding FP8 vector from $B$, expands them to FP16, and combines them using a widening \ac{fma}~(\texttt{vfwmacc.vf}).
\circled{2}{Green}~The E8M0 block scales are loaded from $A_s$ (as a scalar) and $B_s$ (as a vector).
The 8-bit exponents are combined and converted to FP32 using a sequence of integer instructions~\cite{islamoglu2025mxdotp}.
\circled{3}{BrickRed}~The unscaled dot products and the expanded scales are combined using a vector-vector \ac{fma} into the global accumulator vector.
At the end, the result is written back to $C$.

To improve performance, we manually unroll the inner loop to parallelize loads with arithmetic operations.
We also process multiple rows in parallel ($M_\text{tile} = 2$), maximizing data reuse within the \ac{vrf}.
Finally, we implement a similar kernel accumulating in BF16, where the widening \ac{fma} is replaced with a single-width instruction (\texttt{vfmacc.vf}).

\subsection{Analysis}

\begin{figure}[t]
	\centering
	\includegraphics{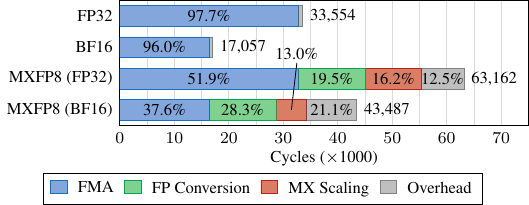}
    \vspace{-1.5em}
	\caption{
        \acs{vau} cycles spent executing different instruction types during \ac{matmul} kernels ($N = 128$).
	}
	\label{fig:baseline_analysis}
    \vspace{-0.5em}
\end{figure}

We evaluate our baseline MXFP8-\ac{matmul} kernels with a $64 \times 64$ output matrix and an inner dimension of $N = 128$ on the Spatz cluster (\Cref{sec:background_spatz}), comparing them to standard FP32 and BF16 \ac{matmul}.
As shown in \Cref{fig:baseline_analysis}, the \ac{mx} kernels have runtimes of 63,162~cycles (FP32 accumulation) and 43,487~cycles~(BF16).
Compared to them, the regular FP32 and BF16 kernels are \qty{88}{\percent} and \qty{155}{\percent} faster, respectively.

To examine the overhead of our \ac{mx} kernels, we analyze the utilization of the functional unit (\acs{vau}) and break down the execution time by instruction type.
In the standard FP32 kernel, \qty{97.7}{\percent} of \ac{vau} cycles are spent on ``useful'' \acp{fma}.
While the MXFP8-to-FP32 kernel requires the same amount of time to perform (widening) \acp{fma}, it performs significant additional work:
\qty{19.5}{\percent} of runtime is used for vector and scalar \ac{fp} conversions, while an additional \qty{16.2}{\percent} is spent converting and applying block scales.
Furthermore, software emulation incurs significant additional overhead of around \qty{12.5}{\percent}, which has multiple reasons.
First, the use of multi-step mixed-precision operations requires frequent \texttt{vtype} changes.
Second, the large number of intermediate results increases register pressure. This requires the use of lower \ac{lmul} values, which in turn reduces the amount of data processed per instruction.
Third, the increased loop nesting leads to more control-flow-related instructions.

The results for BF16 are similar in absolute terms.
However, as BF16 \acp{fma} have higher throughput, the \ac{fma} part of total runtime decreases to \qty{37.6}{\percent} and MX scaling approximately halves while \ac{fp} conversion and overhead cycles stay approximately the same as in the FP32 case.

\subsection{Discussion}

Our analysis of \ac{mx} software emulation reveals significant performance limitations inherent to the current \ac{rvv} \ac{isa}.
While these results were obtained on a specific implementation (i.e., Spatz), the identified bottlenecks, explicit \ac{fp} conversions and software-managed scaling, are fundamental.

Consequently, the software-emulated approach fails to translate the compact representation of \ac{mx} formats into a computational advantage.
This introduces an undesirable trade-off:
while \ac{mx} formats reduce memory footprint and bandwidth, standard \ac{fp} remains the more performant option. To resolve this and unlock the full potential of MX formats, native hardware support for \ac{mx} operations is essential.
\section{The \vmxdotp ISA Extension}

\subsection{Design Goals}

Motivated by the fundamental inefficiencies of software-emulated \ac{mx} operations, we aim to design an \ac{rvv} \ac{isa} extension that enables efficient \ac{mxmatmul} through a native \ac{mxdp} primitive.
Our design is guided by several key goals:
\begin{itemize}
    \item
        To eliminate the overhead of software scaling, the extension should \textbf{apply \ac{mx} scales directly in hardware}.
    \item
        Similar to \ac{fma}, the \ac{mxdp} instructions must include a \textbf{fused accumulation} step.
        This avoids extra \ac{fp} addition and normalization overhead.    
    \item
        The extension should support \textbf{multiple formats}:
        This includes MXFP8\textsubscript{E5M2/E4M3} and MXFP4\textsubscript{E2M1} elements, with accumulation in both the specification-mandated FP32 and the more compact BF16 format.
    \item
        The new instructions should \textbf{integrate with the \ac{rvv} programming model}, be vector-length agnostic, and include both vector-vector and vector-scalar variants.
    \item
        The design must allow \textbf{efficient microarchitectures}, achieving high computational throughput and targeting near-full FPU utilization at low complexity and cost.
    \item
        The extension should support \textbf{flexible block sizes} selected through software, and should not be architecturally constrained to the standard block size of~32.
\end{itemize}

\subsection{Challenges}

\begin{figure}
    \centering
    \includegraphics[]{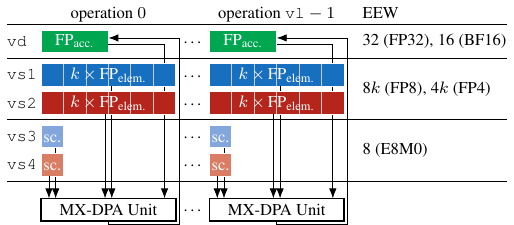}
    \vspace{-0.3em}
    \caption{
        Vector register layout for vector-vector \vmxdotp instructions with block size~$k$.
        There are \texttt{vl} independent \acs{mxdpa} operations.
    }
    \vspace{-0.3em}
    \label{fig:isa_vector_layout}
\end{figure}

Given these design goals, we first consider a comprehensive \acf{mxdpa} instruction that computes a full 32-element dot product between two \ac{mx} blocks, applies the block scales, and accumulates the result.
Its vector data layout is illustrated in \cref{fig:isa_vector_layout} (with $k = 32$).

There are several challenges to address to make this instruction conform to the design goals and feasible to implement:

\begin{enumerate}
    \item The \ac{mxdpa} \textbf{operands vary greatly in bitwidth}, ranging in \ac{eew} from 8~(scales) to 256~bits (32~FP8 elements).
    This diverges significantly from standard \ac{rvv}, where operand widths differ by a factor of two at most. 
    \item The \ac{mxdpa} unit's \textbf{inputs are very wide}, with an \ac{eew} of up to 256~bits for the element operands, compared with at most 32 or 64~bits for existing \ac{rvv} instructions.
    \item There are \textbf{5 source operands} for \ac{mxdpa}, while standard \ac{rvv} instructions are limited to 3 (plus the special mask register).
    This creates challenges for both microarchitectures and instruction encoding.
    \item This operation \textbf{fixes the block size}~$k$, violating the requirement for flexibility in that regard.
\end{enumerate}

Our solution to challenges 1 and~2 is based on a key insight:
An \ac{mxdp} can be decomposed into the sum of multiple smaller dot products that reuse the same block scales.
For example, a 32-wide \ac{mxdp} can be computed by summing the results of four 8-wide dot products.

This insight allows us to reduce the \emph{hardware block size} from 32 to more manageable values.
In particular, we reduce $k$ until \ac{eew} of the \ac{mx} elements equals the scalar \ac{fp} register width (\acuse{flen}\ac{flen}, either 32 or 64).
This avoids introducing new \acp{eew} not present in standard \ac{rvv}.
It also allows holding packed \ac{mx} element operands in scalar \ac{fp} registers, a requirement for vector-scalar instructions.
As described previously, software can implement any \ac{mx} block size which is a multiple of the hardware block size~$k$ (in particular,~32), solving challenge~4.

We do not address challenge~3 at the architectural level.
Rather, as we will show later (\cref{sec:implementation}), it is possible to prefetch and buffer the narrow scale operands with minimal overhead, thereby avoiding the need for expensive additional read ports to the \ac{vrf} banks.

\subsection{Instruction Specification}

\begin{table}[t]
    \centering
    \caption{\vmxdotp instruction variants depending on \ac{flen}}
    \label{tab:vmxdotp_instructions}
    \resizebox{\linewidth}{!}{%
    \begin{tabular}{@{}c@{ ~}c@{ ~}c@{ ~}r@{ ~}r@{ ~}r@{ }r@{ }r@{ }}
        \toprule
        \textbf{\ac{flen}} & \textbf{Instruction} & \textbf{Acc.} & \textbf{\ac{sew}} & \textbf{{EMUL}\textsubscript{elem.}} & \textbf{{EMUL}\textsubscript{sc.}}& $k_\text{FP8}$ & $k_\text{FP4}$ \\
        \midrule
        \multirow{2}{*}{32} & \texttt{vmxdotp.vv/vf} & FP32 & 32 & \emph{LMUL} & $\emph{LMUL} / 4$
            & \multirow{2}{*}{4} & \multirow{2}{*}{8} \\
            & \texttt{vmxdotp.ww/wf} & BF16 & 16 & $2\cdot\emph{LMUL}$ & $\emph{LMUL}/2$ & & \\
        \midrule
        \multirow{2}{*}{64} & \texttt{vmxdotp.ww/wf} & FP32 & 32
            & $2\cdot\emph{LMUL}$ & $\emph{LMUL}/4$ & \multirow{2}{*}{8} & \multirow{2}{*}{16} \\
            & \texttt{vmxdotp.qq/qf} & BF16 & 16 & $4\cdot\emph{LMUL}$ & $\emph{LMUL}/2$ & & \\
        \bottomrule
    \end{tabular}
    }
\end{table}

The different \vmxdotp instructions for $\text{\ac{flen}} = 32$ and $64$ are listed in \cref{tab:vmxdotp_instructions}.
The accumulator precision is set via \ac{sew} (32 for FP32, 16 for BF16), while the element \ac{fp} format (FP8\textsubscript{E5M2/E4M3} or FP4\textsubscript{E2M1}) is selected through a \ac{csr}.
Based on the width ratio of the \ac{flen}-bit element vectors and the \ac{fp} accumulators, the instructions are classified as single-width (ratio~1,~\texttt{v}), narrowing (2,~\texttt{w}), or \emph{quad-narrowing} (4,~\texttt{q}).

For the vector-vector instructions (\texttt{vv}, \texttt{ww}, and \texttt{qq} suffixes), the $i$-th element of the accumulator vector is computed as:
\begin{equation}
    \texttt{vd}[i] \mathrel{+}= \texttt{vs3}[i] \cdot \texttt{vs4}[i] \cdot \sum_{j=0}^{k-1} \texttt{vs1}[ki+j] \cdot \texttt{vs2}[ki+j],
\end{equation}
where \texttt{vs1} and \texttt{vs2} are interpreted in element data format (FP8/FP4), \texttt{vs3} and \texttt{vs4} as E8M0 scales, \texttt{vd} contains the \ac{fp} accumulators, and $k$ is the hardware block size from \cref{tab:vmxdotp_instructions}.
The computation for the vector-scalar instructions (\texttt{vf}, \texttt{wf}, and \texttt{qf} suffixes) is similar, with the first and third operands being broadcast from scalar \ac{fp} registers:
\begin{equation}
    \texttt{vd}[i] \mathrel{+}= \texttt{rs3} \cdot \texttt{vs4}[i] \cdot \sum_{j=0}^{k-1} \texttt{rs1}[j] \cdot \texttt{vs2}[ki+j],
\end{equation}

The required 25~bits to encode the 5 register operands make it infeasible to encode the instructions within the 32-bit encoding space in a standard-compatible way.
There are a number of approaches to reduce the number of bits required to encode the operands, e.g., restricting the number of addressable registers or grouping the scalar \ac{fp} registers into pairs.
However, all such schemes fail to achieve the required reduction in bits without placing severe restrictions on register allocation.

For future standardization, we propose using the longer 48-bit or 64-bit instruction encodings provided by RISC-V~\cite{riscv_unprivilged}, which can easily accommodate 5~full register specifiers.
However, to avoid the complexity of variable-length instruction decoding, prototypes and custom accelerators may recycle unused parts of the 32-bit encoding space instead.
We use this second option for our implementation (\Cref{sec:implementation}).

\subsection{\ac{mxmatmul} Kernel Using \vmxdotp}

We now implement accelerated \ac{rvv} kernels for \ac{mxmatmul} leveraging the new \vmxdotp extension.
Similar to the baseline, we use an outer-product algorithm. However, $B$ is now stored in column-major order, such that elements of the same \ac{mx} block are stored contiguously in memory.
The pseudocode in \Cref{lst:kernel_vmxdotp} illustrates the computation for a single output tile ($1 \times P_\text{tile}$) with MXFP8 inputs and accumulation in FP32.

The code iterates block by block along the reduction dimension (step size $k$).
\circled{1}{RoyalBlue}~In each iteration, elements are loaded from $A$ (packed into a scalar \ac{fp} register) and $B$ (using \ac{flen}-bit strided loads).
\circled{2}{Green}~For each block, scales are loaded from $A_s$ and $B_s$ as before.
As the software block size ($\text{\texttt{BLOCK\_SIZE}} = 32$) differs from the hardware block size ($\text{\texttt{HW\_BLOCK\_SIZE}} = 8$), they are reused across iterations.
\circled{3}{BrickRed}~Finally, the \ac{mxdp} is computed and accumulated using the \texttt{vmxdotp.wf} instruction.

\begin{customlisting}
\begin{lstlisting}[basicstyle=\footnotesize\ttfamily,
    caption={
        \vmxdotp kernel for MXFP8-\ac{matmul} ($1 \times P_\text{tile}$ output tile) with FP32 accumulation.
        We use $\text{\ac{flen}} = 64$, i.e., $\text{\texttt{HW\_BLOCK\_SIZE}} = k_\text{FP8} = 8$.%
        \vspace{1mm}%
    },
    label={lst:kernel_vmxdotp}]
size_t N, P_tile = get_vlmax(SEW_32, LMUL_2);
fp8_t A[1][N];      e8m0_t As[1][N_block];
fp8_t B[P_tile][N]; e8m0_t Bs[N_block][P_tile];
float C[1][P_tile]; double a0, as0;
(*@\textbf{vsetvli}@*)(P_tile, SEW_32, LMUL_M2); v0..1 = (*@\textbf{vmv.v.i}@*)(0);
for (size_t n = 0; n < N; n += HW_BLOCK_SIZE) {
   a0 = A[0][n:n+HW_BLOCK_SIZE];    // 8x FP8 packed
   v4..v7 = (*@\textbf{vlse64.v}@*)(B[:][n]);
   if (n % BLOCK_SIZE == 0){       // once per block
     as0 = As[0][n/BLOCK_SIZE];           // 1x E8M0
     v8 = (*@\textbf{vle8.v}@*)(B[n/BLOCK_SIZE][:]);
   }
   v0..v1 = (*@\textbf{vmxdotp.wf}@*)(v0..v1, a0, v4..v7, as0, v8);
}
c[0][:] = (*@\textbf{vse32.v}@*)(v0..v1);           // store result
\end{lstlisting}
\begin{tikzpicture}[overlay,x=0.7em,y=0.5em]
    \draw[RoyalBlue,thick,Bar-Bar] (1,15.9) -- (1,19.2)
        node[midway,shape=circle,black,inner sep=1pt,draw=none,fill=RoyalBlue,text=white]
        {\small \textbf{1}};
    \draw[Green,thick,Bar-Bar] (1,8.8) -- (1,15.6)
        node[midway,shape=circle,black,inner sep=1pt,draw=none,fill=Green,text=white]
        {\small \textbf{2}};
    \draw[BrickRed,thick] (1,6.7) -- (1,8.5)
        node[midway,shape=circle,black,inner sep=1pt,draw=none,fill=BrickRed,text=white]
        {\small \textbf{3}};
\end{tikzpicture}
\vspace{-1.8em}
\end{customlisting}

As with the baseline, we unroll the loop and process multiple rows in parallel ($M_\text{tile} = 8$) to maximize performance.

To implement MXFP4-\ac{matmul}, only two modifications are required:
we write the relevant \ac{csr} to select FP4 source format, and double \texttt{HW\_BLOCK\_SIZE} to $k_\text{FP4} = 16$.

\section{Hardware Implementation}
\label{sec:implementation}

\begin{figure}
    \centering
    \includegraphics{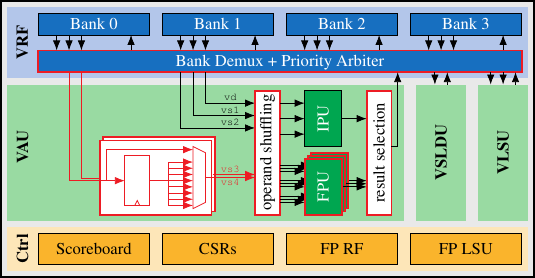}
    \vspace{-1.5em}
    \caption{
        Spatz \ac{vpe} with datapath changes for \vmxdotp integration highlighted in red.
    }
    \label{fig:implementation_spatz}
    \vspace{-1em}
\end{figure}

To evaluate our proposed \vmxdotp extension, we integrate it into the Spatz \ac{vpe}.
Based on Spatz's \ac{flen} of 64, we implement the narrowing (\texttt{w*}) and quad-narrowing (\texttt{q*}) instructions.
Our modifications to Spatz are y in \cref{fig:implementation_spatz}.

For the datapath, we integrate the MXDOTP \ac{fpu}~\cite{islamoglu2025mxdotp}, which includes an 8-wide MXFP8 dot product with FP32 accumulation.
We extend the unit to support 16-wide MXFP4 dot products and BF16 accumulation.
To provide the accumulator and \ac{fp} elements, we reuse the existing infrastructure.
The two \ac{mx} scale operands need to be supplied to the \acp{fpu} separately, which we achieve by adding two read ports (\texttt{vs3}, \texttt{vs4}) to the \ac{vrf}.

The comparatively low read bandwidth required for the scales ($2\times$8~bits per operation) when compared to the elements ($2\times$64~bits) prompts us to fetch a batch of scales at once, buffer them within the \ac{vau}, and consume them progressively over 8~cycles.
This optimization allows us to multiplex the 5~\emph{logical} read ports between \ac{vau} and \ac{vrf} onto the 3~\emph{physical} read ports of each \ac{vrf} bank, avoiding the prohibitive area cost of additional read ports to the memory banks.
In general, this introduces a cycle of overhead every 8~cycles, as the element read requests are stalled during scale prefetching.
However, this overhead is avoided when the operands are mapped to different \ac{vrf} banks, or in the case of vector-scalar instructions (\texttt{vmxdotp.*f}), which only use \texttt{vd}, \texttt{vs2}, and \texttt{vs4}.

We also adjust the operand shuffling to pack the accumulator and scale operands into a single 64-bit value as required by the \acp{fpu}, and modify the result selection to only write 32/16~bits of output per operation in the narrowing/quad-narrowing case.

\section{Evaluation}
\label{sec:evaluation}

\subsection{Physical Implementation}

We implement the baseline and \vmxdotp-enabled Spatz clusters using \textsc{Synopsys Fusion Compiler 2022.03} in \textsc{GlobalFoundries} \qty{12}{\nano\meter} FinFET technology.
We use a target frequency of \qty{0.95}{\giga\hertz} in the worst-case corner (SS, \qty{0.72}{\volt}, \qty{125}{\celsius}).
Our modified cluster successfully meets this target and reaches \qty{1.27}{\giga\hertz} under typical conditions (TT, \qty{0.80}{\volt}, \qty{25}{\celsius}), matching the baseline without introducing a new critical path.

\begin{table}[t]
    \caption{
        Area impact of \vmxdotp at different Spatz hierarchy levels
    }
    \label{tab:evaluation_area}
    \centering
    \begin{tabular}{lrrr}
        \toprule
        \textbf{Hierarchy} & \textbf{Baseline} (kGE) & \textbf{This Work} (kGE) & \textbf{Change} \\
        \midrule
        Cluster & 3995 & 4281 & $+$\phantom{0}\qty{7.2}{\percent} \\
        \midrule
        Core Complex ($\times$2) & 2233 & 2515 & $+$\qty{12.6}{\percent} \\
        \midrule
        \acs{fpu} ($\times$4) & 1264 & 1499 & $+$\qty{18.6}{\percent} \\
        \acs{vau} (w/o~\ac{fpu}/\ac{ipu}) & 74 & 97 & $+$\qty{31.0}{\percent} \\
        \acs{vrf} & 421 & 444 & $+$\phantom{0}\qty{5.5}{\percent} \\
        \bottomrule
    \end{tabular}
\end{table}

\begin{figure*}[t]
    \centering
    \footnotesize
    \begin{tabular}{c@{~ ~}c@{}c}
        \includegraphics{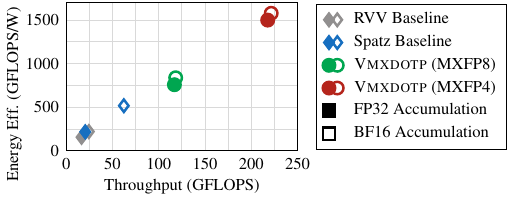} &
        \includegraphics{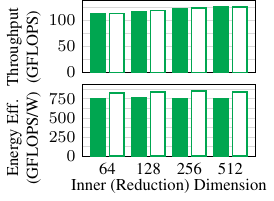} &
        \includegraphics{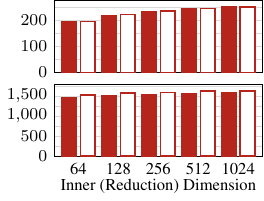} \\[-0.3em]
        \makecell[l]{
            (a) MXFP8 baseline and MXFP8/MXFP4 \vmxdotp kernels\\
            with inner dimension $N = 128$. 
        } &
        \makecell[c]{(b) MXFP8 \vmxdotp kernels.\\~} &
        \makecell[c]{(c) MXFP4 \vmxdotp kernels.\\~}
    \end{tabular}
    \vspace{-0.4em}
    \caption{
        Throughput and energy efficiency for \ac{mxmatmul} kernels with FP32 or BF16 accumulation. 
    }
    \label{fig:evaluation_throughput_efficiency}
    \vspace{-1em}
\end{figure*}

Our \vmxdotp-enabled Spatz cluster has a total area of \qty{4.28}{MGE}, representing an increase of \qty{7.2}{\percent} over the baseline (\qty{12.6}{\percent} at the \ac{cc} level).
A breakdown of the area overhead is presented in \Cref{tab:evaluation_area}.
Most of the increase (\qty{82}{\percent}) is due to the added \ac{mx} dot product unit within the \acp{fpu}, with the remaining overhead split evenly between \ac{vau} and \ac{vrf}.

\subsection{Software Benchmarks}

We evaluate our \vmxdotp \ac{isa} extension on \ac{mxmatmul} with a $64 \times 64$ output matrix, varying inner dimensions, and FP32 or BF16 accumulation.
This is compared with the kernels from \Cref{sec:software_emulation} (\emph{\ac{rvv} baseline}) and enhanced versions using Spatz's custom \emph{MiniFloat-NN} and \emph{ExSdotp} instructions (\emph{Spatz baseline}), both executed on the unmodified Spatz cluster.
All kernels read from and write to the cluster's \qty{128}{\kibi\byte} L1 scratchpad memory.

We use \textsc{Synopsys Prime Time 2022.03} for power estimation under typical conditions at \qty{1}{\giga\hertz}, with switching activities extracted from post-layout simulation.
We average power consumption over five different input samples, which are obtained from DeiT-Tiny~\cite{touvron2021deit} and quantized to MXFP8 and MXFP4 formats using Microsoft’s \emph{Microxcaling} library~\cite{microsoft2024mx}.

\subsection{Throughput and Energy Efficiency}

\Cref{fig:evaluation_throughput_efficiency}a compares our \vmxdotp-accelerated kernels with the \ac{rvv} and Spatz baselines.
Compared with \ac{rvv} emulation, the MXFP8 \vmxdotp kernels achieve a speedup of 7.0$\times$ (FP32 accumulation) and 4.8$\times$ (BF16) at 4.9$\times$ and 3.8$\times$ energy efficiency, respectively.
Results are similar when compared to the FP32 Spatz baseline, while the BF16 Spatz baseline benefits heavily from full support for FP8 arithmetic.
Despite this, our extension still provides a 1.9$\times$~speedup at 1.6$\times$~energy efficiency.
As expected, the MXFP4 \vmxdotp kernels approximately double the throughput and efficiency of their MXFP8 counterparts.
Compared to FP32, using BF16 accumulation with our extension increases energy efficiency by \qtyrange{5}{10}{\percent}, coupled with a small increase in throughput.

Figs.~\ref{fig:evaluation_throughput_efficiency}b and \ref{fig:evaluation_throughput_efficiency}c show the performance of our \vmxdotp extension under various inner dimensions.
For MXFP8, \vmxdotp achieves a throughput of up to 125.0~GFLOPS~(FP32) and 125.4~GFLOPS~(BF16) at an energy efficiency of 753 and 843~GFLOPS/W.
These throughputs correspond to an \ac{fpu} utilization of \qty{97.6}{\percent} and \qty{97.9}{\percent}, respectively.
The results for MXFP4 inputs are similar, achieving a throughput of up to 249.1~GFLOPS (FP32, \qty{97.3}{\percent} utilization) and 250.1~GFLOPS (BF16, \qty{97.7}{\percent} utilization) at an energy efficiency of 1570 and 1632~GFLOPS/W, respectively.

\subsection{Comparison with State of the Art}

We compare \vmxdotp to state-of-the-art \ac{mx} accelerators and a non-\ac{mx} vector processor supporting FP8 arithmetic, as summarized in \cref{tab:soa}.

VEGETA~\cite{nine2025optimizing} and Cuyckens et al.~\cite{cuyckens2025efficient} propose large-scale dataflow accelerators for \ac{matmul} using various \ac{mx} formats.
A direct comparison is challenging, however, as these works target fixed-function accelerators, whereas our design is a fully programmable \ac{vpe} cluster. 
Their system-level figures are extrapolated from \ac{pe}-level synthesis results or simulator estimates, omitting the area, power, and timing overheads of system integration and physical implementation.
In contrast, our cluster-level figures include the interconnect and \qty{128}{\kibi\byte} shared-L1 memory, with energy efficiency results derived from back-annotated post-layout simulations.
Despite this broader scope and full programmability, the energy efficiency of our design remains comparable, achieving 1.8$\times$ that of Cuyckens et al.\ for MXFP8, and 0.94$\times$ for MXFP4.
Unlike VEGETA and Cuyckens et al., which both employ a fixed \ac{mx} block size for quantization, our design supports software-defined block sizes.
This flexibility is crucial given the rapidly evolving landscape of \ac{ai} model quantization and recent work suggesting the use of smaller block sizes for optimal results~\cite{chmiel2025fp4}.

\begin{table}[t]
    \centering
    \vspace{-0.3em}
    \renewcommand{\arraystretch}{1} 
    \setlength{\tabcolsep}{2pt} 
    \caption{Comparison of \vmxdotp with state of the art}
    \label{tab:soa}
    
    \resizebox{\linewidth}{!}{%
    \begin{threeparttable} 

    \begin{tabular}{@{}l@{}c@{~}c@{~}c@{~}c@{~}c@{\hspace{-0.3em}}c@{\hspace{-0.2em}}c@{~}c@{}}
    \toprule
    \textbf{Design} & \textbf{Tech.} & \textbf{Volt.} & \textbf{Freq.} & \textbf{Area}  & \textbf{Input} & \textbf{Accum.} & \textbf{Area Eff.} & \textbf{Energy Eff.} \\
                    & \si{\nano\meter}  & \si{\volt}     & \si{\giga\hertz} & \si{\milli\meter\squared} & \textbf{Format} & \textbf{Format} & \si{\giga{FLOPS}/\milli\meter\squared} & \si{\giga{FLOPS}/\watt} \\
    \midrule
    \multirowcell{3}[0pt][l]{\textbf{VEGETA}\tnote{*†}\\\cite{nine2025optimizing}}
     & \multirow{3}{*}{65} 
     & \multirow{3}{*}{--} 
     & \multirow{3}{*}{0.18} 
     & 1.01 & MXFP8\textsubscript{E5M2} & \multirow{3}{*}{BF16} & 183 & 6460 \\
     &  &  &  & 1.32 & MXFP8\textsubscript{E4M3} &  & 140 & 5680 \\
     &  &  &  & 0.85 & MXFP6\textsubscript{E3M2} &  & 216 & 7912 \\
    \midrule
    \multirowcell{2}[0pt][l]{\textbf{Cuyckens}\\\textbf{et al.}\tnote{*‡}{}\ ~~\cite{cuyckens2025efficient}} & \multirow{2}{*}{16} & \multirow{2}{*}{--} & \multirow{2}{*}{0.40} & \multirow{2}{*}[0.5pt]{8.92} & MXFP8 & \multirow{2}{*}{FP32} & 1469 & 388--420 \\
     & & & & & MXFP4 &  & 2939 & 1667 \\
    \midrule
    \textbf{MXDOTP} \cite{islamoglu2025mxdotp} & 12 & 0.8 & 1.00 & 0.59 & MXFP8 & FP32 & 173 & 356 \\
    \midrule
    \makecell[l]{\textbf{MiniFloat-NN}\\\textbf{Spatz}~\cite{bertaccini2024minifloats}} & 12 & 0.8 & 1.08 & 0.44 & FP8 & FP16 & 307 & 860 \\
    \midrule
    \multirow{2}{*}{\textbf{This Work}} & \multirow{2}{*}{12} & \multirow{2}{*}{0.8} & \multirow{2}{*}{1.00} & \multirow{2}{*}{0.52} & MXFP8 & \multirow{2}{*}{FP32/BF16} & 240/240 & 753/843 \\
     & & & & & MXFP4 &  & 479/481 & 1570/1632 \\
    \bottomrule
    \end{tabular}

    \begin{center}
    \noindent
    \tnote{*}\Ac{pe} level.
    \tnote{†}System-level simulator estimates.
    \tnote{‡}Post-synthesis estimates.
    \end{center}
    \end{threeparttable}
    }
    \vspace{-0.3em}
\end{table}

Turning to programmable, core-based alternatives with instruction extensions, MXDOTP~\cite{islamoglu2025mxdotp} proposes a scalar RISC-V instruction semantically similar to \vmxdotp.
However, its reliance on \acp{ssr} to supply operands represents a significant architectural departure from standard RISC-V.
In contrast, we resolve read port contention microarchitecturally through time-multiplexed \ac{rf} accesses.
Our design is 1.4$\times$ more area-efficient and delivers 2.1$\times$ higher energy efficiency for MXFP8 compared to MXDOTP despite more comprehensive format support.
These results highlight the advantages of vector architectures over scalar processors.

\vmxdotp extends the MiniFloat-NN Spatz~\cite{bertaccini2024minifloats} baseline with MX dot product instructions, trading a small reduction in area and energy efficiency for the superior numerical robustness of \ac{mx} formats compared to scalar minifloats.
The added logic for scale manipulation and multi-operand accumulation accounts for our lower area efficiency and slight decrease (\qtyrange{2}{12}{\percent}) in energy efficiency.
\section{Conclusion}

We presented \vmxdotp, a RISC-V Vector \ac{isa} extension for efficient MXFP8 and MXFP4 dot products, with support for FP32 and BF16 accumulator precisions and software-defined block sizes.
Integrated into Spatz and implemented in a \qty{12}{\nano\meter} technology, \vmxdotp achieves up to 125~MXFP8-GFLOPS at up to 843~MXFP8-GFLOPS/W, and up to 250~MXFP4-GFLOPS at up to 1632~MXFP4-GFLOPS/W.
Compared to software emulation, this represents a speedup of 7.0$\times$ and 4.8$\times$ for FP32 and BF16 accumulation, respectively, while improving energy efficiency by 4.9$\times$ and 3.8$\times$.
These results highlight the need for dedicated block-scaled dot-product-accumulate instructions in \ac{rvv}.

\printbibliography

@report{amd_cdna4,
  title = {Introducing {{AMD CDNA}} 4 {{Architecture}}},
  author = {{Advanced Micro Devices}},
  date = {2025-06},
  url = {https://www.amd.com/content/dam/amd/en/documents/instinct-tech-docs/white-papers/amd-cdna-4-architecture-whitepaper.pdf},
  urldate = {2025-07-25}
}

@article{bertaccini2024minifloats,
  title = {{{MiniFloats}} on {{RISC-V Cores}}: {{ISA Extensions With Mixed-Precision Short Dot Products}}},
  author = {Bertaccini, Luca and Paulin, Gianna and Cavalcante, Matheus and Fischer, Tim and Mach, Stefan and Benini, Luca},
  date = {2024-10},
  journaltitle = {IEEE Transactions on Emerging Topics in Computing},
  volume = {12},
  number = {4},
  pages = {1040--1055}
}

@inproceedings{chmiel2025fp4,
  title = {{{FP4 All}} the {{Way}}: {{Fully Quantized Training}} of {{LLMs}}},
  author = {Chmiel, Brian and Fishman, Maxim and Banner, Ron and Soudry, Daniel},
  booktitle = {{39th Conference on Neural Information Processing Systems (NeurIPS '25)}},
  date = {2025-12}
}

@inproceedings{cuyckens2025efficient,
  title = {Efficient {{Precision-Scalable Hardware}} for {{Microscaling}} ({{MX}}) {{Processing}} in {{Robotics Learning}}},
  booktitle = {2025 {{IEEE}}/{{ACM International Symposium}} on {{Low Power Electronics}} and {{Design}} ({{ISPLED}} '25)},
  author = {Cuyckens, Stef and Yi, Xiaoling and Satya Murthy, Nitish and Fang, Chao and Verhelst, Marian},
  date = {2025-08}
}

@inproceedings{gerogiannis2025deca,
  title = {{DECA: A Near-Core LLM Decompression Accelerator Grounded
on a 3D Roofline Model}},
  shorttitle = {{{DECA}}},
  booktitle = {{58th IEEE/ACM International Symposium on Microarchitecture (MICRO' 25)}},
  author = {Gerogiannis, Gerasimos and Eyerman, Stijn and Georganas, Evangelos and Heirman, Wim and Torrellas, Josep},
  date = {2025-10}
}

@inproceedings{islamoglu2025mxdotp,
  title = {{{MXDOTP}}: {{A RISC-V ISA Extension}} for {{Enabling Microscaling}} ({{MX}}) {{Floating-Point Dot Products}}},
  shorttitle = {{{MXDOTP}}},
  booktitle = {36th {{IEEE International Conference}} on {{Application-specific Systems}}, {{Architectures}} and {{Processors}} ({{ASAP}} '25)},
  author = {İslamoğlu, Gamze and Bertaccini, Luca and Prasad, Arpan Suravi and Conti, Francesco and Garofalo, Angelo and Benini, Luca},
  date = {2025-07}
}

@software{microsoft2024mx,
  title = {{{MX Pytorch Emulation Library}}},
  author = {{Microsoft}},
  date = {2024-08},
  version = {1.1.0},
  url = {https://github.com/microsoft/microxcaling}
}

@inproceedings{nine2025optimizing,
  title = {Optimizing {{Sparse}}/{{Dense VEGETA Accelerator Performance}} with {{Microscaling Quantization}}},
  booktitle = {2025 {{IEEE International Symposium}} on {{Circuits}} and {{Systems}} ({{ISCAS}} '25)},
  author = {Nine, Kazi Barria and Talley, Connor and Mandadi, Ajay Sharma and Krishna, Tushar and Raychowdhury, Arijit},
  date = {2025-05},
}

@online{nvidia_blackwell,
  title = {{{NVIDIA Blackwell Architecture}}},
  author = {{NVIDIA}},
  date = {2025},
  url = {https://www.nvidia.com/en-us/data-center/technologies/blackwell-architecture/},
  urldate = {2025-05-18},
  langid = {american}
}

@standard{ocpmx,
  title = {{{OCP Microscaling Formats}} ({{MX}}) {{Specification}}},
  author = {Rouhani, Bita Darvish and Garegrat, Nitin and Savell, Tom and More, Ankit and Han, Kyung-Nam and Zhao, Ritchie and Hall, Mathew and Klar, Jasmine and Chung, Eric and Yu, Yuan and Schulte, Michael and Wittig, Ralph and Bratt, Ian and Stephens, Nigel and Milanovic, Jelena and Brothers, John and Dubey, Pradeep and Cornea, Marius and Heinecke, Alexander and Rodriguez, Andres and Langhammer, Martin and Deng, Summer and Naumov, Maxim and Micikevicius, Paulius and Siu, Michael and Verrilli, Colin},
  date = {2023-09},
  langid = {english},
  version = {1.0}
}

@article{perotti2025spatz,
  title = {Spatz: {{Clustering Compact RISC-V-Based Vector Units}} to {{Maximize Computing Efficiency}}},
  author = {Perotti, Matteo and Riedel, Samuel and Cavalcante, Matheus and Benini, Luca},
  date = {2025-07},
  journaltitle = {IEEE Transactions on Computer-Aided Design of Integrated Circuits and Systems},
  volume = {44},
  number = {7},
  pages = {2488--2502}
}

@online{qcom_mx,
  author = {Verrilli, Colin},
  organization = {{Qualcomm Technologies}},
  title = {Qualcomm {{Cloud AI}} 100 {{Accelerates Large Language Model Inference}} by \raisebox{0.5ex}{\texttildelow}2x {{Using Microscaling}} ({{Mx}}) {{Formats}}},
  date = {2024-01-09},
  url = {https://www.qualcomm.com/developer/blog/2024/01/qualcomm-cloud-ai-100-accelerates-large-language-model-inference-2x-using-microscaling-mx},
  urldate = {2025-07-28}
}

@standard{riscv_unprivilged,
  title = {The {{RISC-V Instruction Set Manual}}, {{Volume I}}: {{Unprivileged Architecture}}},
  author = {{RISC-V International}},
  date = {2025-05},
  version = {20250508}
}

@online{rouhani2023microscaling,
  title = {Microscaling {{Data Formats}} for {{Deep Learning}}},
  author = {Rouhani, Bita Darvish and Zhao, Ritchie and More, Ankit and Hall, Mathew and Khodamoradi, Alireza and Deng, Summer and Choudhary, Dhruv and Cornea, Marius and Dellinger, Eric and Denolf, Kristof and Dusan, Stosic and Elango, Venmugil and Golub, Maximilian and Heinecke, Alexander and James-Roxby, Phil and Jani, Dharmesh and Kolhe, Gaurav and Langhammer, Martin and Li, Ada and Melnick, Levi and Mesmakhosroshahi, Maral and Rodriguez, Andres and Schulte, Michael and Shafipour, Rasoul and Shao, Lei and Siu, Michael and Dubey, Pradeep and Micikevicius, Paulius and Naumov, Maxim and Verrilli, Colin and Wittig, Ralph and Burger, Doug and Chung, Eric},
  date = {2023-10-19},
  eprint = {2310.10537},
  eprinttype = {arXiv}
}

@inproceedings{rouhani2023shared,
  title = {With {{Shared Microexponents}}, {{A Little Shifting Goes}} a {{Long Way}}},
  booktitle = {50th {{Annual International Symposium}} on {{Computer Architecture}} ({{ISCA}} '23)},
  author = {Rouhani, Bita Darvish and Zhao, Ritchie and Elango, Venmugil and Shafipour, Rasoul and Hall, Mathew and Mesmakhosroshahi, Maral and More, Ankit and Melnick, Levi and Golub, Maximilian and Varatkar, Girish and Shao, Lei and Kolhe, Gaurav and Melts, Dimitry and Klar, Jasmine and L'Heureux, Renee and Perry, Matt and Burger, Doug and Chung, Eric and Deng, Zhaoxia and Naghshineh, Sam and Park, Jongsoo and Naumov, Maxim},
  date = {2023-06}
}

@article{satyamurthy2024optimization,
  title = {Optimization of Block-Scaled Integer {{GeMMs}} for Efficient {{DNN}} Deployment on Scalable in-Order Vector Processors},
  author = {Satya Murthy, Nitish and Catthoor, Francky and Verhelst, Marian},
  date = {2024-09},
  journaltitle = {Journal of Systems Architecture},
  shortjournal = {Journal of Systems Architecture},
  volume = {154},
  pages = {103236}
}

@inproceedings{touvron2021deit,
  title = {Training Data-Efficient Image Transformers \& Distillation through Attention},
  booktitle = {38th {{International Conference}} on {{Machine Learning}} ({{ICML}} '21)},
  author = {Touvron, Hugo and Cord, Matthieu and Douze, Matthijs and Massa, Francisco and Sablayrolles, Alexandre and Jegou, Herve},
  date = {2021-07},
  pages = {10347--10357},
  eventtitle = {International {{Conference}} on {{Machine Learning}}}
}

@online{zvfofp4min,
  title = {{{OFP4}} Conversion Extension {{Zvfofp4min}}, {{Version}} 0.1},
  author = {Waterman, Andrew},
  date = {2025-05-27},
  url = {https://github.com/aswaterman/riscv-misc/blob/e1e20a75c9a9fa797519fcc11ee997c7a7be4503/isa/zvfofp4min.adoc},
  urldate = {2025-05-30}
}

@online{zvfofp8min,
  title = {{{OFP8}} Conversion Extension {{Zvfofp8min}}, {{Version}} 0.2.1},
  author = {Waterman, Andrew},
  date = {2025-05-27},
  url = {https://github.com/aswaterman/riscv-misc/blob/e1e20a75c9a9fa797519fcc11ee997c7a7be4503/isa/zvfofp8min.adoc},
  urldate = {2025-05-30}
}

\end{document}